\documentclass[1p]{elsarticle}
\usepackage{amsmath,amssymb,graphicx}
\begin{document}
\title{A general phase noise relationship for four-wave mixing}


\author{Aravind P. Anthur$^\dagger$, Regan T. Watts$^*$, Tam N. Huynh$^*$, Deepa Venkitesh$^\dagger$ and Liam P. Barry$^*$}
\address{$^\dagger$Department of Electrical Engineering, Indian Institute of Technology Madras, Chennai - 36, India; \\$^*$ Rince Institute, School of Electronic Engineering, Dublin City University, Dublin 9, Ireland.}





%
%

\begin{abstract}
We propose and verify the use of the power spectral density of the FM noise spectrum to study the phase noise
relationship between the four-wave mixing components.
\end{abstract}


%

\maketitle 

\section{Introduction}


Laser phase noise is a characteristic that determines
the quality of the signal, when information is encoded onto both the phase and amplitude of the optical carrier as is the case for advanced modulation formats.
When four-wave mixing (FWM) process is used for all-optical wavelength conversion of advanced modulation formats, it is important to know the 
phase noise transfer from the mixing frequencies to the generated frequencies. R. Hui and J. Zhou have studied
the phase noise increase due to FWM for the partially-degenerate and non-degenerate case in terms of optical linewidth \cite{Hui_1992, Zhou_1994}. 
We have proposed a dual correlated pumping technique that prevents the phase noise increase due to FWM \cite{Aravind_2013_2, Aravind_2013_3}. 
The phase noise of an optical source has frequency dependent and independent contributions \cite{Kikuchi_2012}. These are not 
identifiable in the conventional linewidth measurement techniques \cite{Okoshi_1980}. A more basic quantity is the power spectral density (PSD) of the FM noise spectrum, where the frequency
dependent and independent contributions are clearly distinguishable. 
We propose and verify the phase noise relationship between the FWM components using FM noise instead of linewidth as the metric.
PSD of the FM noise spectrum of different FWM components are compared to the predicted values when one of the mixing frequencies
have very high frequency dependent phase noise. 
The experimental results presented are those for a partially degenerate scheme, but similar relationships holds good for non degenerate 
scheme given in \cite{Zhou_1994} and the dual correlated pumping scheme given in \cite{Aravind_2013_2}.

\section{Theory}
\begin{figure}[h]
\centering
\includegraphics[scale=0.5]{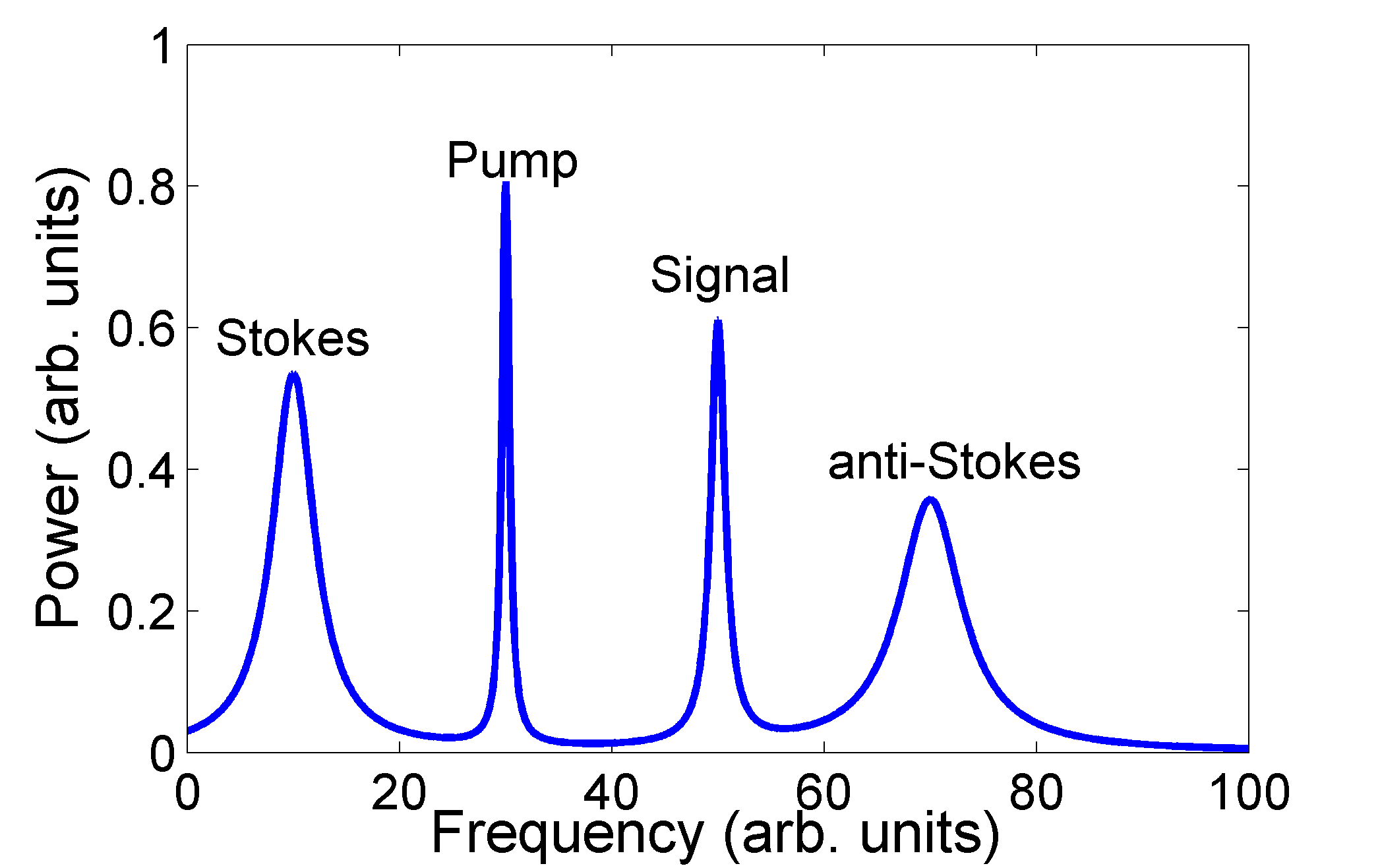}
\caption{Spectral representation of the partially degenerate four-wave mixing scheme.}
\label{Scheme}
\end{figure}

Four-wave mixing (FWM) is a third order nonlinear process where three frequencies ($\omega_{pump-1}$, $\omega_{pump-2}$ and $\omega_{signal}$) mix in a nonlinear medium, generating two more 
frequencies ($\omega_{Stokes}$ at the lower frequency side and $\omega_{anti-Stokes}$ at the higher frequency side). 
When $\omega_{pump-1} = \omega_{pump-2} = \omega_{pump}$, then the process is referred to as
partially-degenerate FWM. The spectral representation of this scheme is given in Fig. \ref{Scheme}.

The relationship for linewidth given in \cite{Hui_1992} and \cite{Zhou_1994} is derived from the phase-error variance
of the FWM components and is true only for the frequency independent (white) phase noise. This linewidth relationship is 
given by \cite{Hui_1992},

\begin{equation}
\Delta \omega_{Stokes} = 4\Delta \omega_{pump} + \Delta \omega_{signal},
\label{lw_Stokes}
\end{equation}

\begin{equation}
 \Delta \omega_{anti-Stokes} = 4\Delta \omega_{signal} + \Delta \omega_{pump}.
\label{lw_anti-Stokes}
\end{equation}

When frequency dependent phase noise is dominant, like in the case of SG-DBR lasers \cite{Kai_2010}, the 
linewidth and phase error variance are not linearly related and hence the relations given in \cite{Hui_1992} and \cite{Zhou_1994}
no longer holds good. The linewidth relationship when the frequency dependent phase noise is dominant is given by \cite{Aravind_2013_3},

\begin{equation}
\Delta \omega_{Stokes}^2 = 4\Delta \omega_{pump}^2 + \Delta \omega_{signal}^2,
\label{lw_Stokes_2}
\end{equation}

\begin{equation}
 \Delta \omega_{anti-Stokes}^2 = 4\Delta \omega_{signal}^2 + \Delta \omega_{pump}^2.
\label{lw_anti-Stokes_2}
\end{equation}

The choice of Eq. \ref{lw_Stokes}/Eq. \ref{lw_anti-Stokes} over Eq. \ref{lw_Stokes_2}/Eq. \ref{lw_anti-Stokes_2} (or vice versa) is not clear
when the proportion of these two types of phase noise is different in the lasers. This becomes 
very important (and difficult to predict) in an optical network where there are multiple wavelength conversions occurring with signal/pump having different
proportions of frequency independent and dependent phase noise. This issue can be resolved if we use a more basic quantity - PSD
of FM noise, for quantifying the phase noise of the FWM components. We find that the relationship derived using the PSD of 
FM noise is independent of the type of phase noise and hence is a general phase noise relationship. 
%
%
%
%
%
%

When $\omega_{pump} < \omega_{signal}$, the FM noise ($S_F(f)$) relationship between the FWM components for the partially-degenerate scheme is given by, 

\begin{equation}
 S_{F}(f)_{Stokes} = 4S_{F}(f)_{pump} + S_{F}(f)_{signal},
\label{Stokes}
\end{equation}

\begin{equation}
 S_{F}(f)_{anti-Stokes} = 4S_{F}(f)_{signal} + S_{F}(f)_{pump}.
\label{anti-Stokes}
\end{equation}

%

The phase error variance and the PSD of the FM noise is related by \cite{Kikuchi_2012},  

\begin{equation}
\sigma_{\Delta \theta}(\tau)^2 = 4 \int_0^{\infty}\left(\frac{\mbox{sin}(\pi f \tau)}{f}\right)^2 S_F(f) df. 
\label{variance}
\end{equation}

Hence the phase error variance relationship and the FM noise relationship between the FWM components should be the same.
Equation \ref{variance} is generally true irrespective of the type of the phase noise. 
In the next section, we experimentally verify that the FM noise relationships given in Eq. \ref{Stokes}  and Eq. \ref{anti-Stokes}
is true at all frequencies ($f$) of the FM noise spectrum and hence independent of the type of phase noise.

\section{Experimental setup}

\begin{figure}[h]
 \centering
\includegraphics[scale=0.45]{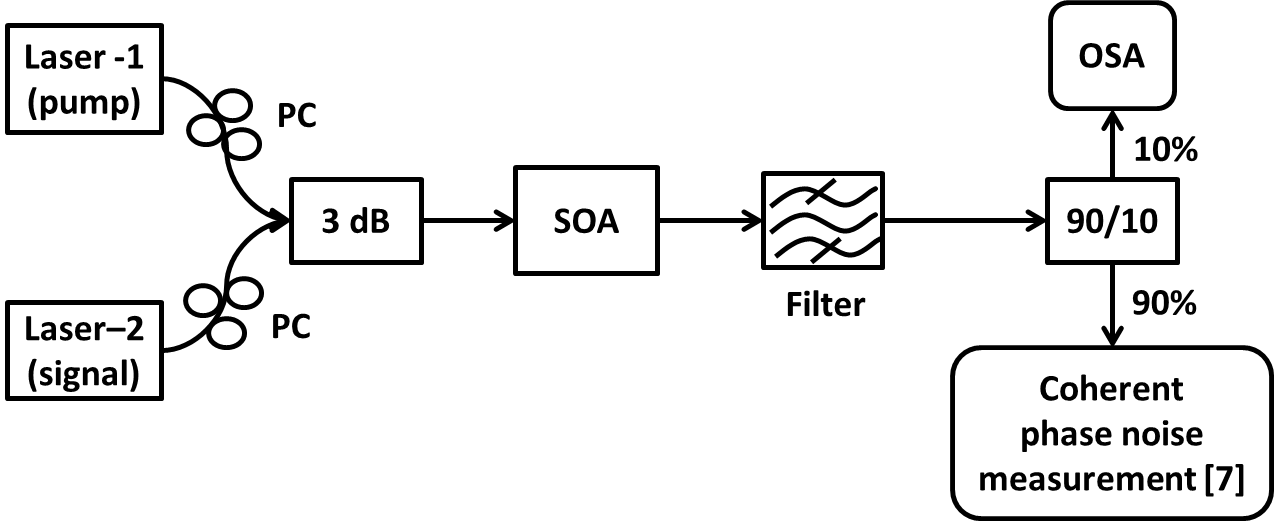}
\caption{Experimental setup for coherent phase noise measurement of four-wave mixing components.}
\label{Experimental_setup}
\end{figure}

Figure \ref{Experimental_setup} shows the schematic of the experimental setup to study the FM  noise relationship 
between the FWM components for the partially-degenerate FWM scheme. Light from the lasers (Laser - 1 and Laser-2) 
are combined using a 3 dB coupler and passed through a SOA where the FWM occurs. Polarization controllers (PC)
are used for best conversion efficiency. The output of the SOA is filtered and 
taken through a 90:10 coupler to analyze the FWM components separately. The 10\% output of the coupler is used for monitoring the spectrum
using an optical spectrum analyzer (OSA). 90\% is used for coherent phase noise measurements \cite{Tam_2012}.

\section{Results}

\begin{figure}[h]
\centering
\includegraphics[scale=0.4]{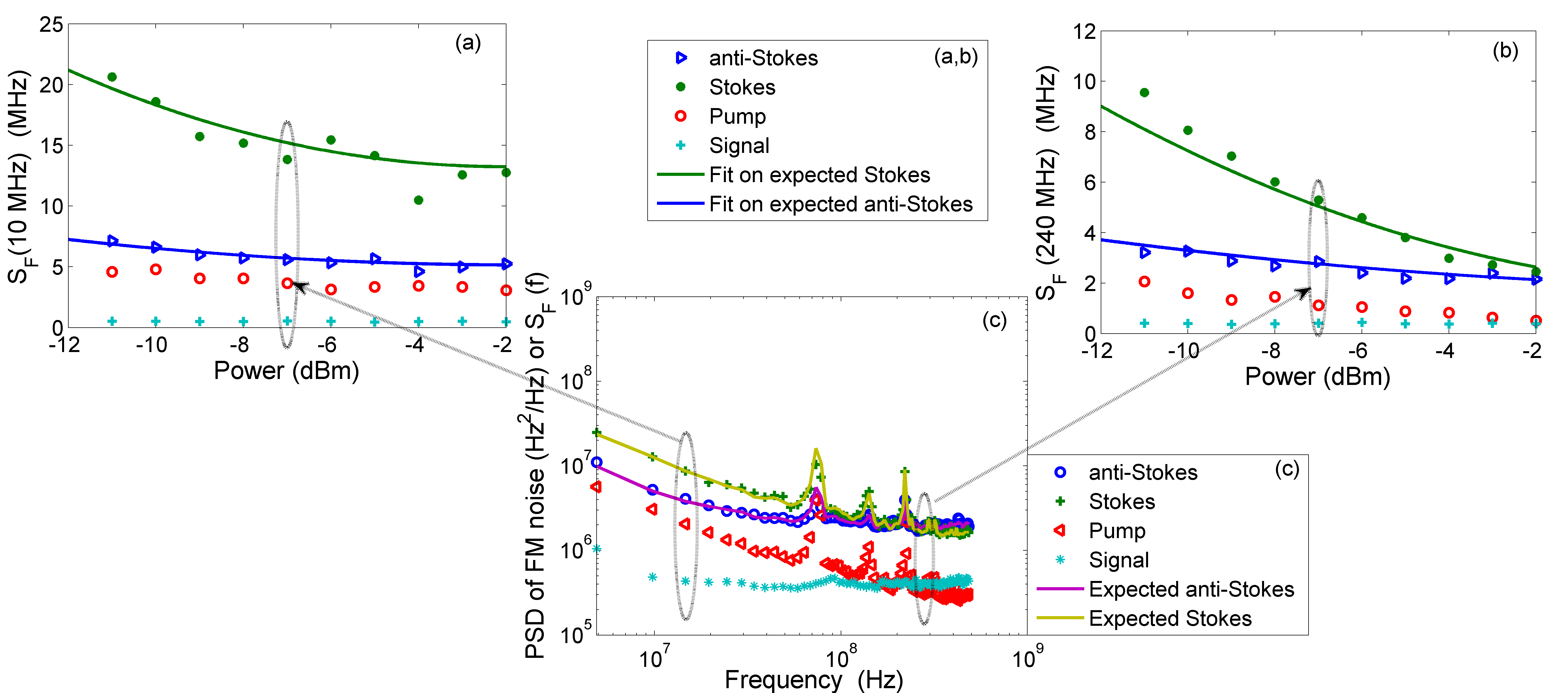}
\caption{(a) Variation of $S_F (f)$ of various FWM components as a function of pump power, in the frequency dependent phase noise region, (b)Variation of $S_F (f)$ of various FWM components as a function of pump power, in the frequency independent phase noise region, (c) Power spectral density (PSD) of the different FWM components for a partially-degenerate four-wave mixing scheme for a signal power of -7 dBm. }
\label{Results}
\end{figure}

Figure \ref{Results} shows the experimental results. Figure \ref{Results}(a) shows the variation of $S_F(f)$ as a function of pump
power at the low frequency (or frequency dependent phase noise) region (at approximatley 10 MHz in Fig. \ref{Results}(c)). Figure \ref{Results}(b) shows the variation of $S_F(f)$ as a function of pump power at high frequency (or frequency independent phase noise) region (at approximately 240 MHz in Fig. \ref{Results}(c)). Continuous lines in Fig. \ref{Results}(a) and Fig. \ref{Results}(b) is obtained by doing a fit on the expected Stokes and expected anti-Stokes $S_F(f)$ values. The expected Stokes and expected anti-Stokes is obtained by substituting the experimental results of $S_F(f)$ of signal and pump in Eq. 5 and Eq. 6 respectively. 
Figure \ref{Results}(c) shows the PSD of the FM noise spectrum of the different FWM components. 
The SG-DBR laser used here as the pump has higher frequency dependent
 phase noise, hence is used to study whether the FM noise relationships given in Eq. \ref{Stokes} and Eq. \ref{anti-Stokes} is dependent/independent 
 of the type of phase noise \cite{Kai_2010}. The continuous lines in Fig. \ref{Results}(c) are obtained using the relationships given in Eq. \ref{Stokes} and Eq. \ref{anti-Stokes}
 and it is clearly seen that expected results follow the experimental results at the low frequency (frequency dependent phase noise)
 region as well as the high frequency (frequency independent phase noise) region.

\section{Conclusions}
The linewidth relationships between the FWM components
detailed till date holds good only in systems
where either the frequency independent or frequency dependent phase noise is dominant, and not otherwise. The linewidth relationships are different
in these two regimes. We propose the use of power spectral density of FM noise
as the metric to study phase noise relationships between the different FWM components, instead of linewidth. We verify that the FM noise relationship between
the FWM components is independent of the type of phase noise and hence is a general phase noise relationship. The proposed relations
holds good even when the proportion of frequency independent and dependent phase noise is unknown. 
The proposed generalized phase noise relations would be more relevant than the linewidth relations in the context of wavelength conversion 
in coherent optical communication systems.


 \section*{Acknowledgements}
%
 The authors acknowledge the Incoming Traveling Fellowship from Science Foundation Ireland (SFI). 
The research leading to these results is also
supported by the People Programme (Marie Curie Actions) of the European Union’s Seventh
Framework Programme FP7/2007-2013/ under REA grant agreement n$^0$318941.

%
%

\end{document}